# String trees

Julius D'souza
cyrrhic@ufl.edu
Department of Electrical and Computer Engineering
University of Florida

**Abstract**

A string-like compact data structure for unlabelled rooted trees is given using $2n$ bits.

## Introduction

Classically, the tree data structure is a dynamic version of the labelled tree in graph theory: each node points to parent or child nodes, and adding or removing nodes from a tree is equivalent to manipulating pointers of appropriate nodes. This structural interpretation of the graph-theoretic tree requires $O(n \log n)$ space since there are $n^{n-2}$ trees with $n$ nodes. The number of labelled trees with $n$ nodes is given by Cayley's formula which can be traced back to Borchardt [1]. Enumerating the number of unlabelled trees, however, has proven to be rather difficult. No closed formulae are presently known, and aside from the discovery of the Redfield-Polya Theorem, little progress on this enumeration problem was made until Otter's work in 1948. Otter[2] provides an asymtotic estimate for unlabelled rooted trees: given $n$ nodes, there are $A\alpha^n n^{-5/2}$ such trees where $A \approx 0.4399$ and $\alpha \approx 2.996$. Hence, the space $S(n)$ needed to represent an $n$-node unlabelled tree is:

$$S(n) = log(A) + n \cdot log(\alpha) - \frac{5}{2} \cdot log(n),$$
$$= O(n).$$

Specifically, $S(n) \approx n \log(\alpha)$. Since $\alpha \approx 2.996$, a per-node representation using an integer alphabet requires $\log \lceil \alpha \rceil = \log(3)$ bits per tree node. The rest of this work describes a data structure using 2 bits per node, or a 4-letter alphabet.

# Tree Linearization

Consider the children $C$ of a tree node $t$ with some total ordering $c_1 \leq c_2 \leq \cdots \leq c_n$, $c_i \in C$. Each $x \in C$ either has children of its own or has none, and $x$ is either the last child of $t$, i.e. $x = \min C$, or is not. These two facts alone suffice in producing a node-based encoding for trees. Consider the following alphabet and semantics:

$x \to$ The node has no children and is not its parent's last child.
$y \to$ The node has children and is not its parent's last child.
$X \to$ The node has no children and is its parent's last child.
$Y \to$ The node has children and is its parent's last child.

For notational purposes and consistency, the root node is ascribed the letter $Y$. The alphabet $\{x, y, X, Y\}$ with the semantics just given completely describes the parent-child relation. Since breadth-first traversal creates a total ordering over the children of each parent node, it is combinatory with the given encoding in capturing the structure of unlabelled trees. For illustration, a level-order walk is performed on the tree $T$ illustrated in breadth-first traversal order in Figure 1 in deriving its equivalent string-tree form; refer to Figure 2 for an illustration of the nodal linearization performed during the walk. Thus follows the walk:

$$\begin{aligned} \text{first level} &\to \text{Y}, \\ \text{second level} &\to \text{xyY}, \\ \text{third level} &\to \text{XyX}, \\ \text{fourth level} &\to \text{Y}, \\ \text{fifth level} &\to \text{xyY}, \\ \text{sixth level} &\to \text{xXxxX}. \end{aligned}$$

Concatenation gives the string tree form of $T$: "YxyYXyXYxyYxXxxX" - or at least the string tree form of $T$ with respect to breadth-first traversal is given. Other forms of ordered tree traversal, such as depth-first traversal, are employable.

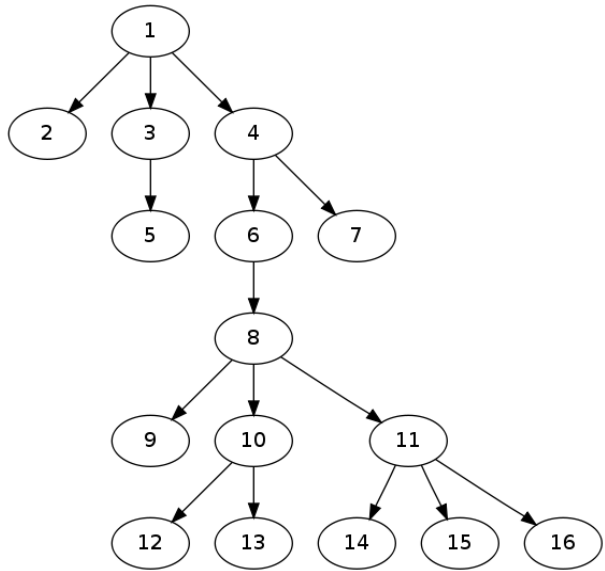

Figure 1: The tree $T$

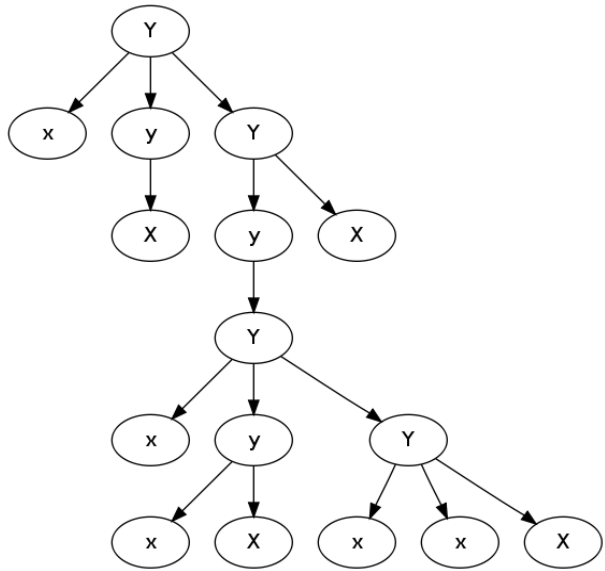

Figure 2: Linearization of the tree $T$

Formalizing the described procedure, the following unlabelled encoding algorithm follows:

Input: a labelled tree $T$
Output: an unlabelled breath-first representation of $T$ in a string $S$

1. Set $S$ to the string "Y".

2. Create a list $L$ of the nodes of $T$ in breadth-first traversal order excluding the root node.

3. Iterate over $L$, adding the following characters to $S$:

    - $x$ if the node has no children and is not the last of its parent's children
    - $y$ if the node has children and is not the last of its parent's children
    - $X$ if the node has no children and is the last of its parent's children
    - $Y$ if the node has children and is the last of its parent's children

4. Return $S$.

Let strings be treated as arrays that start at 0. The following algorithm produces the labelled tree form from an unlabelled representation:

Input: an unlabelled breath-first representation of $T$ in a string $S$ with length $n$
Output: a labelled tree $T$

1. Create a root node for $T$ labelled 0.

2. If $S$ has multiple characters, set the counters $i$ to 1; otherwise, end.

3. Create two empty stacks $A$ and $B$, and push 0 to $A$.

4. While $A$ is not empty:

    (a) Create a node $n$ with the label of the current value of $i$.
    (b) Declare $n$ as the child of the last pushed value in $A$ in the tree $T$.
    (c) If $S[i]$ is "y" or "Y", push $i$ to $B$; if $S[i]$ is "X" or "Y", pop $A$.
    (d) $i \leftarrow i + 1$.

5. Swap $A$ and $B$.

6. If $i$ equals $n$, return $T$; otherwise, return to step 4.

# Concluding Remarks

The string tree provides a tradeoff against its labelled cousins: though most classical tree operations require O($n$) time for the string tree, string operations are readily available for tree manipulation: i.e., concatenation, substring search, and most interestingly string replacement. Furthermore, the plethora of edit distance metrics for strings, i.e. Levenshtein distance, are usable in determining tree similarity.

The string tree alphabet trivially forms a regular language, so regular expressions can be executed directly on a tree structure. As an example, the operation s/(x)*X/X/ kills every node but the last child of a node with no grandchildren on a breadth-first string tree. Regular expression semantics would naturally vary with the type of tree traversal used in forming the string tree.

# Work Cited